\documentclass[12pt]{article}

\usepackage{amsmath}
\usepackage{amssymb}
\usepackage{amsthm}
\usepackage{epsfig}

\newcommand{\prt}{\partial}
\def\Re{{\mathbb R}}

\def\Pr{{\mathbb P}}

\def\B{{\mathcal B}}
\def\D{{\mathcal D}}

\def\be{\begin{equation}}
\def\ee{\end{equation}}

\def\ni{\noindent}

\newtheorem{theorem}{Theorem}

\theoremstyle{definition}

\begin{document}
\bibliographystyle{plain}

\title{{\Large\bf  All the lowest order PDE for spectral gaps of Gaussian matrices}}
\author{Igor Rumanov\footnote{e-mail: igorrumanov@math.ucdavis.edu; current address: MSRI, 17 Gauss Way, Berkeley, CA 94720, e-mail: irumanov@msri.org} \\
{\small Department of Mathematics, UC Davis, 1 Shields Avenue, CA 95616} \\ 
{\small RAS Institute for Structural Macrokinetics and Material Science,} \\
{\small 142432 Chernogolovka, Moscow oblast, Russia}}
%\date{2010}

\maketitle

\bigskip

\begin{abstract}
\par Tracy-Widom (TW) equations for one-matrix unitary ensembles (UE) (equivalent to a particular case of Schlesinger equations for isomonodromic deformations) are rewritten in a general form which allows one to derive all the lowest order equations (PDE) for spectral gap probabilities of UE without intermediate higher-order PDE. This is demonstrated on the example of Gaussian ensemble (GUE) for which all the third order PDE for gap probabilities are obtained explicitly. Moreover, there is a {\it second order} PDE for GUE probabilities in the case of more than one spectral endpoint.
\par This approach allows to derive all PDE at once where possible, while in the method based on Hirota bilinear identities and Virasoro constraints starting with different bilinear identities leads to different subsets of the full set of equations. 
\end{abstract}

\newpage

\section{Introduction: Schlesinger-TW Equations and $\tau$-functions}

\par We consider the unitary ensembles (UE) of Hermitian random matrices. Various large $n$ limits for this case belong to important widespread universality classes: bulk limit leads to equations for the sine kernel~\cite{JMMS, TW1, Me04} whereas edge limit (largest eigenvalues) leads to Airy kernel~\cite{TW-Airy, TW1, Me04}. One studies the operator $K^J$ with the kernel $K^J(x, y) = K(x, y)\chi_{J^c}(y)$, where $K(x, y) = (\varphi(x)\psi(y) - \psi(y)\varphi(x))/(x-y)$ and $\chi_{J^c}(y)$ is equal to zero on $J$ and one on its complement $J^c$ in $\Re$. Important for this theory is the resolvent kernel of $K$ (we will use short-hand notation $K$ for $K^J$), $R(x, y)$, the kernel of $K(I-K)^{-1}$, which is defined by the operator identity

\be
(I + R)(I - K) = I.\ \label{eq:K-R}
\ee

\ni One introduces~\cite{JMMS, IIKS, TW1} the auxiliary functions 

\be
Q(x;J) = (I-K)^{-1}\phi(x), \hspace{1cm} P(x;J) = (I-K)^{-1}\psi(x),\ \label{eq:Q-P}
\ee

\noindent and auxiliary inner products~\cite{TW1}, which are functions of only the endpoints $a_k$ of $J$,

\be
u = (q, \phi\chi_{J^c}), \hspace{1cm} v = (q, \psi\chi_{J^c}) = (p, \phi\chi_{J^c}), \hspace{1cm} w = (p, \psi\chi_{J^c}).\ \label{eq:u-v-w}
\ee

\noindent Then the resolvent kernel $R(x, y)$ is~\cite{JMMS, IIKS, TW1}

\be
R(x, y) = \frac{Q(x;J)P(y;J) - P(x;J)Q(y;J)}{x-y},\ (x,y \in J^c,\ x \neq y)\ \label{eq:R}
\ee

\noindent and (prime denotes the derivative w.r.t.~$x$, e.g.~$Q'(x;J) = dQ(x;J)/dx$)

\be
R(x, x) = P(x;J)Q'(x;J) - Q(x;J)P'(x;J).   \label{eq:Rd}
\ee

\noindent The Tracy-Widom (TW) equations for one-matrix case are the equations for the functions of the endpoints --- $R(a_j, a_k)$,

\be
q_j = Q(a_j;J), \hspace{1cm} p_j = P(a_j;J),   \label{eq:qp}
\ee

\noindent and the auxiliary functions $u$, $v$, $w$. Among them there are universal equations valid for any Hermitian one-matrix unitarily invariant model. They read~\cite{TW1}

\be
\frac{\prt q_j}{\prt a_k} = (-1)^kR(a_j, a_k)q_k,\ \label{eq:qjl}
\ee

\be
\frac{\prt p_j}{\prt a_k} = (-1)^kR(a_j, a_k)p_k,\ \label{eq:pjl}
\ee

\be
R_{jk} \equiv R(a_j, a_k) = \frac{q_jp_k - p_jq_k}{a_j-a_k},\ \label{eq:Rjl}
\ee

\noindent for $j \ne k$ and

\be
(-1)^{j-1}\prt_j\ln\tau^J = R_{jj} \equiv R(a_j, a_j) = p_j\frac{\prt q_j}{\prt a_j} - q_j\frac{\prt p_j}{\prt a_j},\ \label{eq:Rj}
\ee

\be
\frac{\prt u}{\prt a_j} = (-1)^jq_j^2,\ \label{eq:uj}
\ee

\be
\frac{\prt v}{\prt a_j} = (-1)^jq_jp_j,\ \label{eq:vj}
\ee

\be
\frac{\prt w}{\prt a_j} = (-1)^jp_j^2.\ \label{eq:wj}
\ee

\ni By definitions of the matrix integral over $J^n$ (or $J^\infty$), $\tau^J$, and the resolvent operator $R$, we have the fundamental relation between them, expressed by the first equation in formula (\ref{eq:Rj}). This is just a consequence of the defining connection, the equality of two different expressions for the probability of all eigenvalues to lie in $J$,

%\begin{equation}
%R^J(a_j,a_j) = (-1)^{j-1}\frac{\prt\ln\tau^J}{\prt a_j}.  \label{eq:25} %\eqno(25) %\eqno(tau-R)
%\end{equation}

\begin{equation}
\det(I - K^J) = \frac{\tau^J}{\tau},  \label{eq:K-tau} %\eqno(26)
\end{equation}

\noindent where $\tau$ is the corresponding matrix integral over whole $\Re^n$. The integrals $\tau^J$ and $\tau$ can be considered as particular values of certain $\tau$-functions of integrable hierarchies, see e.g.~\cite{ASvM, AvM7, PvM2007}. 
\par The other equations are not universal, their particular form depends on the potential $V(x)$. Let 

\be
q_j' = Q'(a_j; J), \hspace{1cm} p_j' = P'(a_j;J),   \label{eq:q'p'}
\ee

\noindent Then the non-universal equations are %For the finite $n$ Gaussian (Hermite) case they are$^{17}$

\be
\frac{\prt q_j}{\prt a_j} = q_j' - \sum_{k\ne j}(-1)^kR(a_j, a_k)q_k,   \label{eq:qjj}   %-a_jq_j + (\sqrt{2n}-2u)p_j - \sum_{k\ne j}(-1)^kR(a_j, a_k)q_k,\ \label{eq:46)
\ee

\be
\frac{\prt p_j}{\prt a_j} = p_j' - \sum_{k\ne j}(-1)^kR(a_j, a_k)p_k,   \label{eq:pjj}   %a_jp_j - (\sqrt{2n}+2w)q_j - \sum_{k\ne j}(-1)^kR(a_j, a_k)p_k,\ \label{eq:47)
\ee

\noindent along with

\be
R(a_j, a_j) = p_jq_j' - q_jp_j' + \sum_{k\ne j}(-1)^kR(a_j, a_k)(q_jp_k - p_jq_k).   \label{eq:Rjj} %-2a_jq_jp_j + (\sqrt{2n}-2u)p_j^2 + (\sqrt{2n}+2w)q_j^2 + \sum_{k\ne j}(-1)^kR(a_j, a_k)(q_jp_k - p_jq_k),\ \label{eq:48)
\ee

%$$
%\frac{\prt R(a_j, a_j)}{\prt a_j} = -2q_jp_j - \sum_{k\ne j}(-1)^kR(a_j, a_k)^2.\ \label{eq:49)%(4-17)
%$$

%\par It follows from (39)--(41), (46) and (47) that

%$$
%\sum_k\frac{\prt}{\prt a_k}q_jp_j = (\sqrt{2n}-2u)p_j^2 - (\sqrt{2n}+2w)q_j^2,
%$$

%\noindent and the last equation together with (43), (45) gives the first integral obtained in Ref. $17$:

%$$
%2\sum_j(-1)^jq_jp_j = -(2u-\sqrt{2n})(2w+\sqrt{2n}) - 2n,
%$$

%\noindent From (qjk)--(Rjk), (qjj) and (pjj) it follows also that

%\be
%\sum_l\frac{\prt q_j}{\prt a_l} = q_j',    \label{eq:Dq} %-a_jq_j + (\sqrt{2n}-2u)p_j,\ \label{eq:51)
%\ee

%\be
%\sum_l\frac{\prt p_j}{\prt a_l} = p_j'.    \label{eq:Dp} %a_jp_j - (\sqrt{2n}+2w)q_j.\ \label{eq:52)
%\ee

%\noindent One also has general equations for derivatives of $R_{jj}$:

%$$
%\prt_lR_{jj} = \prt_l(p_jq_j' - q_jp_j') - (-1)^lR_{jl}^2,
%$$

%$$
%\prt_jR_{jj} = \prt_j(p_jq_j' - q_jp_j') + \sum_{l \ne j}(-1)^lR_{jl}^2.
%$$

\noindent The above equations are equivalent to the Schlesinger equations involving $2\times 2$ matrices $A_j$,

\begin{equation}
A_j = (-1)^{j-1}\left(\begin{array}{cc} q_jp_j & -q_j^2 \\ p_j^2 & -q_jp_j \end{array}\right) = \frac{\prt}{\prt a_j}\left(\begin{array}{cc} -v & u \\ -w & v \end{array}\right), \label{eq:Schles} 
\end{equation}

\ni see~\cite{HarTW, Pal} and~\cite{UniUE} for details.
\par The paper is organized as follows. In section two we show our general form of equations derived from the above. In section 3 we specify it to the GUE ensemble and obtain explicitly all independent third order PDE for its gap probabilities, for arbitrary number of spectral endpoints. In section 4 we consider the simplest representative example of two endpoints for GUE in more detail, show that some of the previous third order equations are redundant in this case and expose also a second order PDE (lower order than Painlev\'e IV analogs!) satisfied by GUE probabilities. In section 5 we derive the results of section 2 which are the basis for all the rest. In the appendix we demonstrate the application of Adler-Shiota-van Moerbeke approach~\cite{ASvM}--\cite{AvM7} to GUE probabilities based on our system from~\cite{IR1}. The last sections presents the conclusions.

\section{The results. General case}

The system of Schlesinger-TW equations can be reduced in general to the following convenient form, which, at least for the Gaussian case, will immediately give all the 3-rd order PDE in the logarithm of the gap probability, $\ln\Pr$, without intermediate higher-order PDE to integrate.

\begin{theorem}

There are several series of universal equations independent of potential (i.e. of $X_j$, $Y_j$ defined below): the two ``1-point" equations,

\be
\prt_j v\prt_jG_j = G_j\prt^2_{jj}v - \prt_jT\prt_jF,   \label{eq:Tj*}
\ee

\be
G_j^2 = (\prt_jF)^2 + 4F(\prt_jv)^2,   \label{eq:Pj}
\ee 

\ni and the four series of ``2-point" equations,

\be
P_{jl} \equiv \prt_jF\prt_lF - G_jG_l + 4F\prt_jv\prt_lv = 2F(a_j-a_l)^2\prt^2_{jl}T,   \label{eq:Pjl}
\ee

\be
G_l\prt_jF - G_j\prt_lF = 2F(a_j-a_l)\prt^2_{jl}v,   \label{eq:Ajl}
\ee

\be
\prt_jG_l - \prt_lG_j = 2(a_j-a_l)\prt^2_{jl}v,   \label{eq:Ljl}
\ee
  
\be
(a_j-a_l)(\prt_jG_l + \prt_lG_j) = 4(\prt_lv\prt_jF-\prt_jv\prt_lF).   \label{eq:Gjl}
\ee

\ni Also there are two series of non-universal equations where the data dependent on the potential (``spectral curve") enter, let $X_j = q_j'/q_j$, $Y_j = p_j'/p_j$, then   %$X_j = Q'(a_j; J)/q_j$, $Y_j = P'(a_j; J)/p_j$ :

\be
\prt_j\D v = \prt_j v(X_j+Y_j),   \label{eq:Dvj}
\ee

\be
\left(\prt_j\B_0 - a_j\prt_j\D\right)T = \prt_j v(Y_j-X_j),   \label{eq:Tj}
\ee

\noindent Here $\B_k = \sum_j a_j^{k+1}\prt_j$, $\D = \B_{-1} = \sum_j \prt_j$, $T = \ln\tau^J$, definitions of $F$ and $G_j$ are somewhat different for finite size $n$ and infinite matrices. For finite $n$

\be
F = UW \equiv \frac{\tau_{n+1}^J}{\tau_n^J}\cdot\frac{\tau_{n-1}^J}{\tau_n^J},   \label{eq:Fn}
\ee

\be
G_j = W\prt_jU - U\prt_jW,  \label{eq:Gn}
\ee

\be
u \equiv u_n \equiv (\varphi, (I-K_n^J)^{-1}\varphi) = \sqrt\frac{\tau_{n+1}\tau_{n-1}}{(\tau_n)^2}\left(1-\frac{\tau_{n+1}^J/\tau_{n+1}}{\tau_n^J/\tau_n}\right),  \label{eq:uU}
\ee

\be
w \equiv w_n \equiv (\psi, (I-K_n^J)^{-1}\psi) = \sqrt\frac{\tau_{n+1}\tau_{n-1}}{(\tau_n)^2}\left(\frac{\tau_{n-1}^J/\tau_{n-1}}{\tau_n^J/\tau_n}-1\right),  \label{eq:wW}
\ee
 
\be
\left. v \equiv v_n \equiv (\varphi, (I-K_n^J)^{-1}\psi) \equiv (\psi, (I-K_n^J)^{-1}\varphi) = -\frac{\prt}{\prt t_1}\ln\frac{\tau_n^J}{\tau_n}\right|_{t=0}. \label{eq:v-tau}
\ee

\ni For infinite ensembles

\be
F = -uw,   \label{eq:Finf}
\ee

\be
G_j = u\prt_jw - w\prt_ju. \label{eq:Ginf}
\ee

\end{theorem}

\par We observe that we have 4 series of 1-point equations (depending on one subscript $j$) and 4 series of 2-point equations (depending on two subscripts, $j$ and $l$). This means that for the case of $N$ spectral endpoints the total number of equations here is $4N + 4N(N-1)/2 = 2N(N+1)$. The model-dependent functions are supposed to be ultimately expressed in terms of the variables $T$, $v$, $F$ and $\{G_j, j = 1,\dots,N\}$. Then the number of dependent variables in the equations is going to be equal to $N + 3$. We get only an {\it upper bound} on the number of independent PDE for the single function $T$ only, assuming that the other $N+2$ variables are eliminated, reducing the number of equations by $N+2$: 

$$
\# \text{ equations for } T \le 2N(N+1) - (N+2) = 2N^2 + N - 2.
$$

\ni As we will see below on the example of GUE, the actual number of independent equations for $T$ is less than this bound. It seems difficult to determine the exact number in general. 
\par Multiplying the equations (\ref{eq:Dvj}) and (\ref{eq:Tj}) by $a_j^k$ with an integer $k \ge 0$ and summing over the endpoints $a_j$, one gets

\begin{theorem}

The model-specific equations for one-matrix UE can be presented in the following general form in terms of operators $\{\B_k, k \ge -1\}$ of Virasoro subalgebra:

\be
\B_{k-1}\D v = \Phi_k,   \label{eq:hvk}
\ee

\be
(\B_{k-1}\B_0 - \B_k\D)T = \Gamma_k,   \label{eq:hTk}
\ee

\noindent where we denoted $\Phi_k = \sum_j a_j^k \prt_j v(X_j+Y_j)$ and $\Gamma_k = \sum_j a_j^k \prt_j v(Y_j-X_j)$. 

\end{theorem}

\ni Multiplying the universal ``2-point" equations by $a_j^ka_l^m$, then summing over $j$ and $l$ and introducing quantities

$$
\hat G_k = \sum_j a_j^k G_j
$$

\noindent leads to 

\begin{theorem}

The four common series of equations for all one-matrix UE in terms of operators $\{\B_k, k \ge -1\}$ of Virasoro subalgebra read:

\be
\B_{k-1}F\B_{m-1}F - \hat G_k\hat G_m + 4F\B_{k-1}v\B_{m-1}v = 2F(\B_{k+1}\B_{m-1} + \B_{k-1}\B_{m+1} - 2\B_k\B_m)T,   \label{eq:hPkm}
\ee 

\be
\hat G_m\B_{k-1}F - \hat G_k\B_{m-1}F = 2F(\B_k\B_{m-1} - \B_{k-1}\B_m + \B_{k+m-1})v,   \label{eq:hAkm}
\ee 

\be
\B_{k-1}\hat G_m - \B_{m-1}\hat G_k + (k-m)\hat G_{k+m-1} = 2(\B_k\B_{m-1} - \B_{k-1}\B_m + \B_{k+m-1})v,   \label{eq:hLkm}
\ee 

\be
\B_{m-1}\hat G_{k+1} - \B_m\hat G_k - \hat G_{k+m} = 2(\B_{m-1}v\B_{k-1}F - \B_{k-1}v\B_{m-1}F) - (\B_{k+1}\B_{m-1} + \B_{k-1}\B_{m+1} - 2\B_k\B_m)v.   \label{eq:hGkm}
\ee

\ni All the other equations that UE satisfy in terms of the operators $\B_k$ only, are combinations of these equations and their derivatives.

\end{theorem} 

%\bigskip
%\noindent (?)Obviously, for $N$ endpoints only the first $N$ $\B$-operators starting with $\D \equiv \B_{-1}$ are linearly independent, and so algebraically independent are equations (hvk), (hTk) with $0 \le k \le N-1$ and equations (hPkm), (hAkm), (hLkm) and (hGkm) with $0 \le k, m \le N-1$. So the total is $2N^2 + 2N$ (since permutation $k \leftrightarrow m$ doesn't change anything) independent such equations.
\par It is worth to write out explicitly the components $k, 0$ and $k, k$ of the ``multivariable Painlev\'e-type" series  (\ref{eq:hPkm}), which contains indeed all multivariable generalizations of Painlev\'e equations:

\be
\B_{k-1}F\D F - \hat G_k\hat G_0 + 4F\B_{k-1}v\D v = 2F(\B_{k+1}\D + \B_{k-1}\B_1 - 2\B_k\B_0)T,   \label{eq:hPk0}
\ee 

\be
(\B_{k-1}F)^2 - \hat G_k^2 + 4F(\B_{k-1}v)^2 = 2F(\B_{k+1}\B_{k-1} + \B_{k-1}\B_{k+1} - 2(\B_k)^2)T.   \label{eq:hPkk}
\ee 

%$$
%\hat G_m\B_{k-1}F - \hat G_k\B_{m-1}F = 2F(\B_k\B_{m-1} - \B_{k-1}\B_m + \B_{k+m-1})v,   \label{eq:\hat A_{km})
%$$ 

%$$
%\B_{k-1}\hat G_m - \B_{m-1}\hat G_k + (k-m)\hat G_{k+m-1} = 2(\B_k\B_{m-1} - \B_{k-1}\B_m + \B_{k+m-1})v,   \label{eq:\hat L_{km})
%$$ 

%$$
%\B_{m-1}\hat G_{k+1} - \B_m\hat G_k - \hat G_{k+m} = 2(\B_{m-1}v\B_{k-1}F - \B_{k-1}v\B_{m-1}F) - (\B_{k+1}\B_{m-1} + \B_{k-1}\B_{m+1} - 2\B_k\B_m)v.   \label{eq:\hat G_{km})
%\ee 

\bigskip

\par One can eliminate $G_j$-variables using equations (\ref{eq:Pj}), (\ref{eq:Pjl}) and (\ref{eq:Ajl}) and obtain 2 independent series of universal PDE containing only $T$ and functions $v$ and $F$, which, for finite size ensembles, are related to $T$ as~\cite{UniUE} $v = -\prt_t T|_{t=0}$ and $F = \prt^2_{tt}T|_{t=0}$, where $t$ is the first time of one-dimensional Toda lattice hierarchy~\cite{AvM7, IR1}. They can be written as e.g.

\begin{theorem}

For any two spectral endpoints $a_j$ and $a_l$, the following two PDE hold for all one-matrix UE:

\be
(\prt^2_{jl}v)^2 - (a_j-a_l)^2(\prt^2_{jl}T)^2 + 4\prt^2_{jl}T\prt_jv\prt_lv = 0,     \label{eq:Tvjl}
\ee

\be
(a_j-a_l)^2(F(\prt^2_{jl}v)^2 - \prt^2_{jl}T\prt_jF\prt_lF) - (\prt_lv\prt_jF - \prt_jv\prt_lF)^2 = 0.     \label{eq:TvFjl}
\ee

\ni All the other such equations that UE satisfy are combinations of these and their derivatives.

\end{theorem}

{\bf Remark.} In the previous paper we used a higher-order PDE\footnote{we know it is higher-order due to the connection with Toda lattice since in general $F = \prt^2_{tt}T|_{t=0}$, where $t$ is the first Toda time.},

\be
\prt_jv(\prt^2_{jj}F + 2(\prt_jv)^2) = \prt_jF\prt^2_{jj}v - \prt_jT G_j,   \label{eq:Fjj*}
\ee

\noindent which is in fact the eq.~(\ref{eq:Tj*}) where $\prt_jG_j$ has been eliminated using the derivative of eq.~(\ref{eq:Pj}) w.r.t.~$a_j$. Then from eqs.~(\ref{eq:Pj}) and (\ref{eq:Fjj*}) one obtains the higher-order universal ``1-point" PDE of our previous paper~\cite{UniUE},

\be
\left(\prt_jv(\prt^2_{jj}F + 2(\prt_jv)^2) - \prt_jF\prt^2_{jj}v \right)^2 = (\prt_jT)^2((\prt_jF)^2 + 4F(\prt_jv)^2),   \label{eq:T1j}
\ee

\noindent while the eqs.~(\ref{eq:Tvjl}) and (\ref{eq:TvFjl}) are ``2-point" equations.

\section{Results for the Gaussian case}

For the Gaussian case the expressions for the non-universal quantities $X_j$ and $Y_j$ can be found~\cite{TW1} and written as

$$
\prt_j v(X_j+Y_j) = 2\prt_jF,
$$

$$
\prt_j v(Y_j-X_j) = 2(G_j + a_j\prt_jv),
$$

\noindent so the two non-universal equations (\ref{eq:Dvj}) and (\ref{eq:Tj}) give, respectively,

$$
\D v = 2(F-n/2),
$$

$$
2G_j = \left(\prt_j\B_0 - a_j\prt_j\D\right)T - 2a_j\prt_jv.
$$

\noindent Then the linear equation (\ref{eq:Ljl}) together with the last two relations result in formulas

\be
2v = \D T,   \label{eq:v}
\ee

\be
4F = \D^2 T + 2n,   \label{eq:F}
\ee

\be
2G_j = \left(\prt_j\B_0 - 2a_j\prt_j\D\right)T.   \label{eq:Gj}
\ee

\noindent Thus, all auxiliary variables are expressed in terms of $T$. Substituting these expressions into the other universal equations one finds the following two theorems:

%$$
%\left(\prt_j\B_0 - 2a_j\prt_j\D\right)T \prt^2_{jj}\D T - \prt_j\D T\prt_j\left(\prt_j\B_0 - 2a_j\prt_j\D\right)T = \prt_j T\prt_j\D^2T.   \label{eq:T_j*)
%$$

\begin{theorem}

The logarithm $T$ of a spectral gap probability for GUE satisfies the following ``1-point" 3rd order equations:

\be
\prt_j\B_0 T \prt^2_{jj}\D T - \prt_j\D T\prt^2_{jj}\B_0 T = \prt_j T\prt_j\D^2T - 2(\prt_j\D T)^2.   \label{eq:Tj*g}
\ee

\be
(\prt_j\D^2T)^2 - 4(\left(\prt_j\B_0 - 2a_j\prt_j\D\right)T)^2 + 4(\D^2 T + 2n)(\prt_j\D T)^2 = 0.   \label{eq:Pjg}
\ee

\ni All the other such equations it satisfies are combinations of these and their derivatives.

\end{theorem}

\noindent Eq.~(\ref{eq:Tj*g}), however, becomes trivial in the case of one endpoint, while in the case of two endpoints the two such equations are a consequence of the other ones, see the next section. Whether this is the case in general, remains unclear at the moment.

\begin{theorem}

The logarithm $T$ of a spectral gap probability for GUE satisfies the following ``2-point" 3rd order equations:

$$
\prt_j\D^2T\prt_l\D^2T - 4\left(\prt_j\B_0 - 2a_j\prt_j\D\right)T\left(\prt_l\B_0 - 2a_l\prt_l\D\right)T + 4(\D^2 T + 2n)\prt_j\D T\prt_l\D T = 
$$

\be
= 8(a_j-a_l)^2(\D^2 T + 2n)\prt^2_{jl}T,   \label{eq:Pjlg}
\ee

\be
\left(\prt_l\B_0 - 2a_l\prt_l\D\right)T \prt_j\D^2T - \left(\prt_j\B_0 - 2a_j\prt_j\D\right)T\prt_l\D^2T = 2(a_j-a_l)(\D^2 T + 2n)\prt^2_{jl}\D T,   \label{eq:Ajlg}
\ee

\be
2(a_j-a_l)(\prt^2_{jl}\B_0 - (a_j+a_l)\prt^2_{jl}\D)T = \prt_l\D T\prt_j\D^2 T - \prt_j\D T\prt_l\D^2 T.    \label{eq:Gjlg}
\ee

\ni All the other such equations it satisfies are combinations of these and their derivatives.

\end{theorem}

\ni We can now count the total number of third order PDE for the logarithm of GUE gap probabilities. We got two 1-point series and three 2-point series of equations, therefore in the case of N endpoints $a_j$ one has $2N + 3N(N-1)/2 = N(3N+1)/2$ independent PDE. This should be reduced by $1$ for $N=1$ since then (\ref{eq:Tj*g}) is trivial and we are left with (\ref{eq:Pjg}) only.
\par The last three series of PDE have their counterparts in terms of $\B$-operators only. Since for the case of GUE we have

\be
\Phi_k = 2\B_{k-1}F = \B_{k-1}\D v = \B_{k-1}\D^2T/2,    \label{eq:hvkg}
\ee

\be
\Gamma_k = 2(\hat G_k + \B_k v) = 2\hat G_k + \B_k\D T = (\B_{k-1}\B_0-\B_k\D)T,     \label{eq:hTkg}
\ee

\be
2\hat G_k = (\B_{k-1}\B_0-2\B_k\D)T,     \label{eq:hGk}
\ee

\ni we obtain immediately, either from theorem 3 or from theorem 6,

\begin{theorem}

The logarithm $T$ of a spectral gap probability for GUE satisfies the following equations in terms of operators $\{\B_k, k \ge -1\}$ of Virasoro subalgebra:

$$
\B_{k-1}\D^2T\B_{m-1}\D^2T - 4(\B_{k-1}\B_0-2B_k\D)T(\B_{m-1}\B_0-2B_m\D)T + 4(\D^2T + 2n)B_{k-1}\D T\B_{m-1}\D T = 
$$

\be
= 8(\D^2T+2n)(\B_{k+1}\B_{m-1} + \B_{k-1}\B_{m+1} - 2\B_k\B_m)T,     \label{eq:hPkmg}
\ee

$$
(\B_{m-1}\B_0-2\B_m\D)T\B_{k-1}\D^2T - (\B_{k-1}\B_0-2\B_k\D)T\B_{m-1}\D^2T = 
$$

\be
= 2(\D^2T+2n)(\B_k\B_{m-1} - \B_{k-1}\B_m + \B_{k+m-1})\D T,     \label{eq:hAkmg}
\ee

$$
\left(\B_{m-1}(\B_k\B_0-2\B_{k+1}\D) - \B_m(\B_{k-1}\B_0-2\B_k\D) + \B_k(\B_{m-1}\B_0-2\B_m\D) -  \right. 
$$

\be
\left. - \B_{k-1}(\B_m\B_0-2\B_{m+1}\D)\right)T = \B_{m-1}\D T\B_{k-1}\D^2T - \B_{k-1}\D T\B_{m-1}\D^2T,    \label{eq:hGkmg}
\ee

\ni All the other equations in terms of the operators $\B_k$ it satisfies are combinations of these and their derivatives.

\end{theorem}

\noindent In fact, eq.~(\ref{eq:Pj}) can be considered as eq.~(\ref{eq:Pjl}) for $a_j = a_l$, therefore, summing (\ref{eq:Pjl}) over both subscripts $j$ and $l$, we get the component in a sense corresponding to (\ref{eq:Pj}) also, namely, for GUE -- the ``diagonal" equation (\ref{eq:hPkmg}) with $m = k$. The total number of algebraically independent equations in the three series is thus $3N(N-1)/2 + N = N(3N-1)/2$ in the case of $N$ endpoints. It is less than the total number of 1-point and 2-point equations above by $N$, as should be since equations (\ref{eq:Tj*g}) do not have a counterpart here.
\par E.g.~the simplest members of the three series are, respectively,

\be
(\D^3T)^2 - 4(\D T - \B_0\D T)^2 + 4(\D^2T + 2n)(\D T)^2 = 16(\D^2T+2n)(\B_1\D - (\B_0)^2 + \B_0)T,     \label{eq:hP00}
\ee

\be
(\D T - \B_0\D T)\B_0\D^2T - ((\B_0)^2T - 2\B_1\D T)\D^3T = 2(\D^2T+2n)(\B_1\D - (\B_0)^2 + \B_0)\D T,     \label{eq:hA10}
\ee

\be
\left(\D(\B_1\B_0-2\B_2\D) - 2\B_0((\B_0)^2-2\B_1\D) + \B_1(\D-\B_0\D)\right)T = \D^2 T\B_0\D^2T - \B_0\D T\D^3T.    \label{eq:hG10}
\ee

\ni Equation (\ref{eq:hP00}) was first found by comparison of two 4th order PDE -- the ``boundary-KP" equation of~\cite{ASvM} and another one derived by the author from ``boundary-Toda-AKNS" system of~\cite{IR1}, see Appendix. Now the general procedure to obtain {\it all the lowest order equations at once}, wherever possible, is outlined.

\section{Example: Gaussian case with two endpoints}

We start with general equations of theorem 1. However, let us introduce notations which will be convenient for Gaussian two-endpoint case. If the two endpoints are $a_j$ and $a_l$, then let

$$
\xi_+ = a_j + a_l, \ \ \ \xi_- = a_j - a_l, \ \ \ \D = \prt_j + \prt_l, \ \ \ \D_- = \prt_j - \prt_l,
$$

$$
G_+ = G_j + G_l, \ \ \ G_- = G_j - G_l.
$$

\ni Then we have

$$
\prt_j = 1/2(\D + \D_-), \ \ \ \prt_l = 1/2(\D - \D_-), \ \ \ G_j = 1/2(G_+ + G_-), \ \ \ G_l = 1/2(G_+ - G_-).
$$

\ni In the new notation equations of theorem 1 read as follows:

$$
(G_+ + G_-)^2 = (\D F + \D_-F)^2 + 4F(\D v + \D_-v)^2, \eqno(\ref{eq:Pj})
$$

\be
(G_+ + G_-)^2 = (\D F + \D_-F)^2 + 4F(\D v + \D_-v)^2, \label{eq:Pl}
\ee

$$
(\D F)^2 - (\D_-F)^2 - G_+^2 + G_-^2 + 4F((\D v)^2 - (\D_-v)^2)) = 2F\xi_-^2(\D^2T - \D_-^2T),  \eqno(\ref{eq:Pjl})
$$

$$
G_+\D_-F - G_-\D F = \xi_-F(\D^2v - \D_-^2v),  \eqno(\ref{eq:Ajl})
$$

$$
\D_-G_+ - \D G_- = \xi_-(\D^2v - \D_-^2v),  \eqno(\ref{eq:Ljl})
$$

$$
\xi_-(\D G_+ - \D_-G_-) = 4(\D v\D_-F - \D_-v\D F),  \eqno(\ref{eq:Gjl})
$$

$$
(\D v + \D_-v)(\D + \D_-)(G_+ + G_-) = (G_+ + G_-)(\D + \D_-)^2v - 2(\D T + \D_-T)(\D F + \D_-F),   \eqno(\ref{eq:Tj*})
$$

\be
(\D v - \D_-v)(\D - \D_-)(G_+ - G_-) = (G_+ - G_-)(\D - \D_-)^2v - 2(\D T - \D_-T)(\D F - \D_-F).   \label{eq:Tl*}
\ee

\ni Adding and subtracting the first two equations above, one gets, respectively,

\be
G_+^2 + G_-^2 = (\D F)^2 + (\D_-F)^2 + 4F((\D v + (\D_-v)^2),  \label{eq:P+}
\ee

\be
G_+G_- = \D F\D_-F + 4F\D v\D_-v.   \label{eq:Px}
\ee

\ni Adding and subtracting (\ref{eq:Pjl}) from (\ref{eq:P+}) gives, respectively,

\be
G_+^2 = (\D F)^2 + 4F(\D v)^2 - \xi_-^2F(\D^2 - \D_-^2)T,  \label{eq:P-r}
\ee

\be
G_-^2 = (\D_- F)^2 + 4F(\D_- v)^2 + \xi_-^2F(\D^2 - \D_-^2)T.  \label{eq:Ps}
\ee

\ni The non-universal equations for the Gaussian two-endpoint case give

\be
4G_+ = 2\D T - \xi_+\D^2T - \xi_-\D\D_-T, \label{eq:G_+}
\ee

\be
4G_- = 2\D_- T - \xi_-\D^2T - \xi_+\D\D_-T - \xi_-(\D^2T - \D_-^2T), \label{eq:G-}
\ee

\ni and already known expressions for $v$ and $F$ in terms of $T$. One sees again that (\ref{eq:Ljl}) is satisfied. Denoting also

$$
r = \D T, \ \ \ S = \D\D_-T, \ \ \ r_- = \D_-T, \ \ \ A = \D^2T - \D_-^2T,
$$

\ni one finds expressions

$$
v = r/2, \ \ \ F = 1/4(\D r + 2n), 
$$

$$
\D r - \D_-r = A, \ \ \ \D r_- = \D_-r = S,
$$

$$
\D v = \D r/2, \ \ \ \D_-v = S/2, \ \ \ (\D^2 - \D_-^2)v = 1/2\D A,
$$

$$
\D_-S = \D^2r - \D A, \ \ \ \D_-^2v = 1/2(\D r - \D A),
$$

$$
\D F = 1/4\D^2r, \ \ \ \D_-F = 1/4\D S,
$$

$$
4G_+ = 2r - \xi_+\D r - \xi_-S, \ \ \ 4G_- = 2r_- - \xi_-\D r - \xi_+S - \xi_-A,
$$

$$
4\D G_+ = -\xi_+\D^2r - \xi_-\D S, \ \ \ 4\D G_- = -\xi_-\D^2r - \xi_+\D S - \xi_-\D A,
$$

$$
4\D_- G_+ = -\xi_+\D S - \xi_-(\D^2r - \D A), \ \ \ 4\D_- G_- = -\xi_+(\D^2r - \D A) - \xi_-\D S - \xi_-\D_- A - 4A.
$$

\ni Using all these expressions, we can rewrite our system of PDE as

$$
(2r - \xi_+\D r - \xi_-S)^2 = (\D^2r)^2 + 4(\D r)^2(\D r + 2n) - 4\xi_-^2(\D r + 2n)A, \eqno(\ref{eq:P-r})
$$

$$
(2r_- - \xi_-\D r - \xi_+S - \xi_-A)^2 = (\D S)^2 + 4S^2(\D r + 2n) + 4\xi_-^2(\D r + 2n)A, \eqno(\ref{eq:Ps})
$$

$$
(2r - \xi_+\D r - \xi_-S)(2r_- - \xi_-\D r - \xi_+S - \xi_-A) = \D^2r\D S + 4\D rS(\D r + 2n), \eqno(\ref{eq:Px})
$$

$$
(2r - \xi_+\D r - \xi_-S)\D S - (2r_- - \xi_-\D r - \xi_+S - \xi_-A)\D^2r = 2\xi_-(\D r + 2n)\D A, \eqno(\ref{eq:Ajl})
$$

$$
2(S\D^2r - \D r\D S) = \xi_-(\xi_+\D A - \xi_-\D_-A - 4A), \eqno(\ref{eq:Gjl})
$$

\ni and equations (\ref{eq:Tj*}) and (\ref{eq:Tl*}) reduce after a long tedious calculation to the following two equations,

\be
2\xi_-A\D^2r = (\xi_-(\D r + A) - 2r_-)\D A + S(\xi_-\D_-A + 4A), \label{eq:rA}
\ee

\be
2\xi_-A\D S = (\xi_-S - 2r)\D A + \D r(\xi_-\D_-A + 4A). \label{eq:SA}
\ee

\ni Let us denote also 

$$
\hat F = \D r + 2n, \ \ \ D_- = \xi_-\D_-A + 4A.
$$

\ni Then, expressing $\D A$ from (\ref{eq:Ajl}) 

$$
2\xi_-\hat F\D A = \hat G_+\D S - \hat G_-\D^2r, \eqno(\ref{eq:Ajl})
$$

\ni and then $D_-$ -- from (\ref{eq:Gjl}),

$$
2\xi_-\hat FD_- = (\xi_+\hat G_+ + 4\hat F\D r)\D S - (\xi_+\hat G_- + 4\hat FS)\D^2r, \eqno(\ref{eq:Gjl})
$$

\ni   Substituting these expressions into (\ref{eq:rA}) and (\ref{eq:SA}), one sees that the last equations become trivial. Thus, they are also redundant, i.e.~are consequences of the rest of the system.
\bigskip
\par {\bf Remark.} One can notice a lot of similarity with the final system of PDE for joint largest eigenvalue probabilities of {\it two coupled} GUE, see the Corollary in section 6 of~\cite{IR3}. This is to be expected in light of the representation of the matrix model with one finite size spectral gap in terms of models of two coupled matrices in~\cite{BoDaEy}, p. 6745. More, however, remains to be understood about these relations.
\bigskip
\par We are left with five independent PDE involving four independent third derivatives of $T$. This means that there should be one {\it second order PDE in T}, i.e. equation of {\it lower} order than multidimensional analogs of Painlev\'e IV. Indeed, this equation can be easily obtained, if we express $(\D^2r\D S)^2$ from eq.~(\ref{eq:Px}) and compare with its expression obtained from the corresponding product of (\ref{eq:P-r}) and (\ref{eq:Ps}). It reads as

\be
(\D r\hat G_-  -  S\hat G_+)^2 = \xi_-^2A((\hat G_-)^2 - (\hat G_+)^2 + 4\hat F((\D r)^2 - S^2 - \xi_-^2A)).   \label{eq:2G}
\ee

\ni It becomes trivial upon the identification of the two endpoints as it should be. The analogous equation for GUE can be obtained also in general, for any number of endpoints. This follows by expressing $(\prt_jF\prt_lF)^2$ from (\ref{eq:Pjl})  and comparing with its expression from the corresponding product of (\ref{eq:Pj}) and (\ref{eq:Pl}), now for any pair of endpoints $a_j$ and $a_l$. Thus, we obtain

\begin{theorem}
 
The logarithm of a GUE gap probability $T$ satisfies a second order PDE,

$$
(\prt_l\D T(\prt_j\B_0T - 2a_j\prt_j\D T) - \prt_j\D T(\prt_l\B_0T - 2a_l\prt_l\D T))^2 = 
$$

\be
4(a_j - a_l)^2\prt^2_{jl}T(-(\prt_j\B_0T - 2a_j\prt_j\D T)(\prt_l\B_0T - 2a_l\prt_l\D T) + (\D^2T + 2n)(\prt_j\D T\prt_l\D T - (a_j - a_l)^2\prt^2_{jl}T)).  \label{eq:2-T}
\ee

\end{theorem}

\section{Transformation of the TW-Schlesinger system}

Here mostly the derivations of the statements in the previous sections are gathered. 
\par An important identity is obvious from definitions in section 1:

$$
(\prt_j v)^2 = \prt_j u\prt_j w.
$$

\ni Taking the second derivatives of $u$, $v$ and $w$ and using the equations for derivatives of $q_j$ and $p_j$, one can obtain:

$$
\prt_{jl}^2u = 2\frac{\prt_ju\prt_lv - \prt_lu\prt_jv}{a_j-a_l},    
$$

$$
\prt_{jl}^2v = \frac{\prt_ju\prt_lw - \prt_lu\prt_jw}{a_j-a_l},
$$

$$
\prt_{jl}^2w = 2\frac{\prt_jv\prt_lw - \prt_lv\prt_jw}{a_j-a_l},
$$

$$
\prt_{jj}^2u = 2(-1)^jq_jq_j' - \sum_{l\ne j}\prt_{jl}^2u,      % = 2\prt_juX_j - \sum_{l\ne j}\prt_{jl}^2u
$$

$$
\prt_{jj}^2v = (-1)^j(p_jq_j' + q_jp_j') - \sum_{l\ne j}\prt_{jl}^2v,    % = \prt_jv(X_j + Y_j) - \sum_{l\ne j}\prt_{jl}^2v
$$

$$
\prt_{jj}^2w = 2(-1)^jp_jp_j' - \sum_{l\ne j}\prt_{jl}^2w.       % = 2\prt_jwY_j - \sum_{l\ne j}\prt_{jl}^2w
$$

\ni The fisrt derivatives of $T$ can be written in two different ways: on the one hand, by (\ref{eq:Rj}), (\ref{eq:qjj}) and (\ref{eq:pjj}),

$$
\prt_jT = (-1)^{j-1}R_{jj} = (-1)^{j-1}(p_jq_j' - q_jp_j') - \sum_{l\ne j}(-1)^{j+l}R_{jl}^2(a_j - a_l),     % = \prt_jv(Y_j - X_j) - \sum_{l\ne j}(-1)^{j+l}R_{jl}^2(a_j - a_l)
$$

\ni on the other hand, by (\ref{eq:Rj}) and (\ref{eq:uj})--(\ref{eq:wj}),

$$
\prt_jT = (-1)^{j-1}R_{jj} = (-1)^{j-1}(p_j\prt_jq_j - q_j\prt_jp_j) = \frac{\prt_ju\prt_{jj}^2w - \prt_jw\prt_{jj}^2u}{2\prt_jv} = -\frac{\prt_jv}{2}\prt_j\ln\left | \frac{\prt_ju}{\prt_jw} \right |.
$$

\ni The terms in the sum of the first formula can be also represented as

$$
(-1)^{j+l}R_{jl}^2(a_j - a_l) = \frac{\prt_ju\prt_lw + \prt_lu\prt_jw - 2\prt_jv\prt_lv}{a_j-a_l}.
$$

\ni  Switching to the variables $U$, $W$, $X_j$, $Y_j$ introduced above, one can rewrite the equations as follows:

$$
\prt_ju\prt_lw = -\prt_jU\prt_lW, \ \ \ (\prt_jv)^2 = -\prt_jU\prt_jW,
$$

$$
\prt_{jl}^2U = 2\frac{\prt_jU\prt_lv - \prt_lU\prt_jv}{a_j-a_l},    
$$

$$
\prt_{jl}^2v = \frac{\prt_jW\prt_lU - \prt_lU\prt_jW}{a_j-a_l},
$$

$$
\prt_{jl}^2W = 2\frac{\prt_jv\prt_lW - \prt_lv\prt_jW}{a_j-a_l},
$$

$$
\prt_{jj}^2U = 2\prt_jUX_j - \sum_{l\ne j}\prt_{jl}^2U,      
$$

$$
\prt_{jj}^2v = \prt_jv(X_j + Y_j) - \sum_{l\ne j}\prt_{jl}^2v,    
$$

$$
\prt_{jj}^2W = 2\prt_jWY_j - \sum_{l\ne j}\prt_{jl}^2W,       
$$

$$
\prt_jT = \prt_jv(Y_j - X_j) - \sum_{l\ne j}(-1)^{j+l}R_{jl}^2(a_j - a_l),
$$

$$
\prt_jT = \frac{\prt_jW\prt_{jj}^2U - \prt_jU\prt_{jj}^2W}{2\prt_jv} = \frac{\prt_jv}{2}\prt_j\ln\left | \frac{\prt_jU}{\prt_jW} \right |.
$$

\ni Now we express everything in terms of new functions $F$ and $G_j$ defined in the statement of theorem 1. The first simple identity becomes

$$
(\prt_jv)^2 = -\frac{(\prt_jF)^2 - G_j^2}{4F},
$$

\ni which is nothing but the equation (\ref{eq:Pj}). Mixed second derivative equations now read as

$$
\prt_{jl}^2U = U\frac{(\prt_jF + G_j)\prt_lv - (\prt_lF + G_l)\prt_jv}{(a_j-a_l)F},    
$$

$$
\prt_{jl}^2v = \frac{G_l\prt_jF - G_j\prt_lF}{2(a_j-a_l)F},     \eqno(\ref{eq:Ajl})   %\label{eq:Ajl)
$$

$$
\prt_{jl}^2W = W\frac{\prt_jv(\prt_lF - G_l) - \prt_lv(\prt_jF - G_j)}{(a_j-a_l)F}.
$$

\ni Since

$$
W\prt_{jl}^2U - U\prt_{jl}^2W = \frac{1}{2}(\prt_jG_l + \prt_lG_j), \ \ \ W\prt_{jl}^2U + U\prt_{jl}^2W = \prt_{jl}^2F - \frac{\prt_jF\prt_lF - G_jG_l}{2F},
$$

\ni it follows that

$$
\prt_jG_l + \prt_lG_j = 4\frac{\prt_jF\prt_lv - \prt_lF\prt_jv}{a_j-a_l},    \eqno(\ref{eq:Gjl})   %\label{eq:Gjl)
$$

\be
\prt_{jl}^2F - \frac{\prt_jF\prt_lF - G_jG_l}{2F} = 2\frac{G_j\prt_lv - G_l\prt_jv}{a_j-a_l}.   \label{eq:Fjl}
\ee

\ni By definitions, either (\ref{eq:Fn}) and (\ref{eq:Gn}) or (\ref{eq:Finf}) and (\ref{eq:Ginf}), there are also equations of Adler-van Moerbeke type~\cite{AvM1}, 

\be
\prt_j\left(\frac{G_l}{F}\right) = \prt_l\left(\frac{G_j}{F}\right),   \label{eq:AvM}
\ee

\ni see~\cite{IR3} for the discussion of their analog for coupled GUE joint distributions first derived in~\cite{AvM1}. The last equation together with (\ref{eq:Ajl}) means that a general {\it linear} PDE holds:

$$
\prt_jG_l - \prt_lG_j = 2(a_j-a_l)\prt^2_{jl}v.   \eqno(\ref{eq:Ljl})  %\label{eq:Ljl)
$$

\ni The diagonal equations then read:

\be
\prt_jG_j = (\prt_jF + G_j)X_j - (\prt_jF - G_j)Y_j - 2\sum_{l\ne j}\frac{\prt_jF\prt_lv - \prt_lF\prt_jv}{a_j-a_l},    \label{eq:Gjj}
\ee

\be
\prt_{jj}^2F - \frac{(\prt_jF)^2 - G_j^2}{2F} = (\prt_jF + G_j)X_j + (\prt_jF - G_j)Y_j - 2\sum_{l\ne j}\frac{G_j\prt_lv - G_l\prt_jv}{a_j-a_l},    \label{eq:Fjj}
\ee

$$
\prt_{jj}^2v = \prt_jv(X_j + Y_j) - \sum_{l\ne j}\frac{G_l\prt_jF - G_j\prt_lF}{2(a_j-a_l)F}.    
$$

\ni Using the above equation (\ref{eq:Ajl}) and the diagonal equation for $\prt_{jj}^2v$, the second derivatives $\prt_{jl}^2v$ can be summed over all $l$, giving

%$$
%\D\prt_jU = 2\prt_jUX_j,      
%$$

$$
\D\prt_jv = \prt_jv(X_j + Y_j).   \eqno(\ref{eq:Dvj})    
$$

%$$
%\D\prt_jW = 2\prt_jWY_j,       
%$$

\ni For $\prt_jT$ one has expressions
 
\be
\prt_jT = \prt_jv(Y_j - X_j) - \sum_{l\ne j}(-1)^{j+l}R_{jl}^2(a_j - a_l) = \prt_jv(Y_j - X_j) + \sum_{l\ne j}\frac{\prt_jF\prt_lF - G_jG_l + 4F\prt_jv\prt_lv}{2(a_j-a_l)F},    \label{eq:Tj1}
\ee

\ni and

$$
\prt_jT = \frac{2F\prt_jF\prt_jG_j - G_j(2F\prt_{jj}^2F - (\prt_jF)^2 + G_j^2)}{8F^2\prt_jv} = \frac{\prt_jF\prt_jG_j - G_j(\prt_{jj}^2F + 2(\prt_jv)^2)}{4F\prt_jv}.
$$

\ni If we sum over $j$ the first of these two equations, the last sum on its right-hand side cancels out since its terms are antisymmetric in indices $j$ and $l$, so we obtain

\be
\D T = \sum_j\prt_jv(Y_j - X_j).   \label{eq:DT}
\ee

\ni The second derivatives of $F$, entering eqs.~(\ref{eq:Fjl}) and (\ref{eq:Fjj}), can in fact be removed from consideration. This is because we can take the corresponding derivatives of equation (\ref{eq:Pj}) and compare with (\ref{eq:Fjl}) and (\ref{eq:Fjj}). The result is that equation (\ref{eq:Fjl}) is redundant -- it follows from the derivative of (\ref{eq:Pj}) w.r.t. $a_l$ and the other equations. As for the equation (\ref{eq:Fjj}), its comparison with the derivative of (\ref{eq:Pj}) w.r.t. $a_j$,

$$
2\prt_jF\prt_{jj}^2F - 2G_j\prt_jG_j + 4\prt_jF(\prt_jv)^2 + 8F\prt_jv\prt_{jj}^2v = 0,
$$

\ni gives immediately the equation (\ref{eq:Tj*}), which is thus also proved. As we will see, at least for Gaussian case using eq.~(\ref{eq:Tj*}) is preferable since it encodes a third order PDE for $T$, while (\ref{eq:Fjj}) corresponds to a fourth order PDE in $T$. So we keep (\ref{eq:Tj*}) and then (\ref{eq:Fjj}) is also redundant.
\par Thus, what remains to prove theorem 1 is to derive equations (\ref{eq:Pjl}) and (\ref{eq:Tj}).

\begin{proof}
\par The simplest way to prove (\ref{eq:Pjl}) and (\ref{eq:Tj}) is to differentiate (\ref{eq:Rj}) over $a_l$, $l \ne j$, and then use (\ref{eq:qjl}), (\ref{eq:pjl}), which gives

$$
\prt^2_{jl}T = (-1)^j(q_j\prt^2_{jl}p_j - p_j\prt^2_{jl}q_j + \prt_lq_j\prt_jp_j - \prt_lp_j\prt_jq_j) = 
$$

$$
= (-1)^{j+l}(q_j\prt_j(R_{jl}p_l) - p_j\prt_j(R_{jl}q_l) + R_{jl}(q_l\prt_jp_j - p_l\prt_jq_j)),
$$

\ni but (\ref{eq:qjl}) and (\ref{eq:pjl}) also imply that

$$
q_j\prt_jp_l - p_j\prt_jq_l = 0,
$$

\ni and therefore, differentiating (\ref{eq:Rjl}), we get

$$
\prt_jR_{jl} = -\frac{R_{jl}}{a_j - a_l} + \frac{p_l\prt_jq_j - q_l\prt_jp_j}{a_j - a_l}.
$$

\ni  Applying the last two equations results in the fundamental relation

\be
\prt^2_{jl}T = -(-1)^{j+l}R_{jl}^2,     \label{eq:TjlR}
\ee

\ni which is equivalent to (\ref{eq:Pjl}) if we express $R_{jl}$ in terms of $F$, $G_j$, $G_l$ and $v$. Now, substituting $\prt^2_{jl}T$ into the summand of (\ref{eq:Tj1}) and gathering terms with derivatives of $T$ together, one obtains equation (\ref{eq:Tj}).
\end{proof}

It remains to prove theorem 4, since all the other results are consequences of theorem 1.

\begin{proof}

We first multiply eq.~(\ref{eq:Pjl}) by $\prt_jF\prt_lF$, take square of eq.~(\ref{eq:Ajl}) and then compare them, getting

$$
2G_j\prt_lF\cdot G_l\prt_jF = 2(\prt_jF\prt_lF)^2 + 8F\prt_jv\prt_lF\cdot \prt_lv\prt_jF - 4F(a_j-a_l)^2\prt_{jl}^2T\prt_jF\prt_lF =
$$

$$
= G_j^2(\prt_lF)^2 + G_l^2(\prt_jF)^2 - 4F^2(a_j-a_l)^2(\prt^2_{jl}v)^2.
$$

\ni Then we use eq.~(\ref{eq:Pj}) to express $G_j^2$ and $G_l^2$, after which some terms cancel out and the last equality above becomes eq.~(\ref{eq:TvFjl}) of theorem 4.
\par To derive eq.~(\ref{eq:Tvjl}), we first write down a direct consequence of (\ref{eq:Pjl}),

$$
(\prt_jF\prt_lF - G_jG_l)^2 = 4F^2((a_j-a_l)^2\prt^2_{jl}T - 2\prt_jv\prt_lv)^2,
$$

\ni then use the square of (\ref{eq:Ajl}) again, obtaining

$$
(\prt_jF\prt_lF)^2 - G_j^2(\prt_lF)^2 - G_l^2(\prt_jF)^2 + 4F^2(a_j-a_l)^2(\prt^2_{jl}v)^2 + G_j^2G_l^2 = 4F^2((a_j-a_l)^2\prt^2_{jl}T - 2\prt_jv\prt_lv)^2.
$$

\ni Applying again eq.~(\ref{eq:Pj}) to eliminate $G_j^2$ and $G_l^2$ from the last equation leads to eq.~(\ref{eq:Tvjl}) after cancellation of several terms.
\par It is clear that the number of such independent equations is two since we used four equations algebraically eliminating two variables -- $G_j$ and $G_l$ from them.

\end{proof}

\section{Conclusions}

All independent lowest order PDE for GUE spectral gap probabilities are explicitly obtained. This gives also an example of explicit integrable hierarchies in terms of isomonodromic ``times'' -- the spectral endpoint variables.

\bigskip

{\bf\large Acknowledgements} \\
Author is grateful to C.A.Tracy for stimulating comments and general encouragement. 
\par This work was done with NSF support under grant DMS-0906387.

\section{Appendix: PDE for Gaussian matrices from bilinear identities and Virasoro constraints}

In \cite{IR1} from Toda-AKNS system we obtained the following system of three equations for Gaussian one-matrix gap probabilities:

\be
\B_{-1}^2\ln \tau_n^J = 2(2U_nW_n - n), \hspace{3cm} \label{eq:0}%(0)
\ee

\be
\B_{-1}^2 U_n = (-2\B_0 + 4n)U_n - 8U_n^2W_n, \hspace{3cm} \label{eq:+}%(+)
\ee

\be
\B_{-1}^2 W_n = (2\B_0 + 4n)W_n - 8U_nW_n^2, \hspace{3cm} \label{eq:-}%(-)
\ee

\noindent where $U_n = \tau_{n+1}^J/\tau_n^J$, $W_n = \tau_{n-1}^J/\tau_n^J$, and the operators $\B_k$ contain partial derivatives w.r.t. the boundary points of the intervals:

$$
\B_k\tau_n^J = \sum_{i \in \{endpoints\}} a_i^{k+1}\frac{\partial}{\partial a_i}\tau_n^J.
$$

It can be solved in general. Let $F = UW \equiv U_nW_n$, $G = W\B_{-1}U - U\B_{-1}W$, $G_0 = W\B_0U - U\B_0W$, $T = \ln\tau_n^J$. We will need the commutation relation:

\be
\B_0\B_{-1} = \B_{-1}(\B_0 - 1).   \label{eq:com}
\ee

\noindent The combination $W\cdot(\ref{eq:+})-U\cdot(\ref{eq:-})$ together with (\ref{eq:0}) gives

$$
\B_{-1}G = -2\B_0F = -\frac{1}{2}\B_0\B_{-1}^2 T = -\frac{1}{2}\B_{-1}(\B_0-1)\B_{-1}T,
$$

\noindent or, taking into account vanishing of all the derivatives at infinity,

\be
G = -\frac{1}{2}(\B_0-1)\B_{-1}T.  \label{eq:G}
\ee
 
\noindent The combination $W\cdot(\ref{eq:+})+U\cdot(\ref{eq:-})$ and the above definitions of new variables lead to

\be
\B_{-1}^2 F - \frac{(\B_{-1}F)^2 - G^2}{2F} = -2G_0 - 16F(F-\frac{n}{2}).   \label{eq:FF}
\ee

\noindent Another combination, $\B_{-1}W\cdot(\ref{eq:+})+\B_{-1}U\cdot(\ref{eq:-})$, results in

\be
\B_{-1}\left(\frac{(\B_{-1}F)^2 - G^2}{4F}\right) = -2(\B_{-1}W \B_0U - \B_{-1}U \B_0W) - 8(F-\frac{n}{2})\B_{-1}F.   \label{eq:dF}
\ee

\noindent If we try to use the expression

$$
2(\B_{-1}W \B_0U - \B_{-1}U \B_0W) = \frac{\B_{-1}FG_0 - \B_0FG}{F} = \frac{\B_{-1}FG_0 +1/2G\B_{-1}G}{F}
$$

\noindent on the right-hand side of (\ref{eq:dF}), we get just eq.~(\ref{eq:FF}) back again and nothing new. One needs to use another expression for the right-hand side of (\ref{eq:dF}):

$$
2(\B_{-1}W \B_0U - \B_{-1}U \B_0W) = \B_{-1}G_0 - (\B_0+1)G,
$$

\noindent which can be obtained by moving the differentiation operators in $\B_{-1}W \B_0U - \B_{-1}U \B_0W$ to the left in two possible ways and then matching them and using eq.~(\ref{eq:com}). Then eq.~(\ref{eq:dF}) becomes

$$
2\B_{-1}G_0 = 2(\B_0+1)G - 8\B_{-1}(F-\frac{n}{2})^2 - \B_{-1}\left(\frac{(\B_{-1}F)^2 - G^2}{2F}\right).   \eqno(\ref{eq:dF})
$$

\noindent Luckily, here $(\B_0+1)G$ can also be expressed as $\B_{-1}$-derivative, using eqs.~(\ref{eq:G}) and (\ref{eq:com}):

$$
2(\B_0+1)G = -(\B_0^2 - 1)\B_{-1}T = -\B_{-1}((\B_0-1)^2-1)T = \B_{-1}(2\B_0 - \B_0^2)T,
$$

\noindent so eq.~(\ref{eq:dF}) is a total derivative and its integration gives

\be
2G_0 = (2\B_0 - \B_0^2)T - 8(F-\frac{n}{2})^2 - \frac{(\B_{-1}F)^2 - G^2}{2F}.   \label{eq:G0}
\ee

\noindent Finally, substituting eq.~(\ref{eq:G0}) into eq.~(\ref{eq:FF}) and using eqs.~(\ref{eq:0}) and (\ref{eq:G}), we obtain a single equation for $T = \ln\tau_n^J$:

\be
(\B_{-1}^4 - 4\B_0^2 + 8\B_0 + 8n\B_{-1}^2)T + 2(\B_{-1}^2T)^2 - \frac{(\B_{-1}^3T)^2 - 4((\B_0-1)\B_{-1}T)^2}{\B_{-1}^2T + 2n} = 0.   \label{eq:bT}
\ee

\noindent This is a fourth order PDE as is the ``boundary-KP" equation of Adler, Shiota and van Moerbeke~\cite{ASvM},

\be
(\B_{-1}^4 + 8n\B_{-1}^2 + 12\B_{0}^2 + 24\B_{0} - 16\B_{-1}\B_{1})\ln\tau_n^J + 6(\B_{-1}^2T)^2 = 0,   \label{eq:bKP}
\ee

\noindent derived from the KP equation. Comparing the eqs.~(\ref{eq:bT}) and (\ref{eq:bKP}), we obtain a {\it third order} equation for $T$. Using another commutation relation:

\be
\B_{-1}\B_1 = \B_1\B_{-1} + 2\B_0,   \label{eq:com1}
\ee

\noindent it can be written as

\be
(\B_{-1}^3T)^2 - 4((\B_0-1)\B_{-1}T)^2 + 4(\B_{-1}^2T + 2n)((\B_{-1}^2T)^2 + (\B_0^2 - \B_0 - \B_1\B_{-1})T) = 0.   \label{eq:Pr}
\ee

\noindent This last equation is a direct multidimensional analog of the well-known Painlev\'e IV equation for single endpoint Gaussian case~\cite{TW1, ASvM}:

\be
(r'')^2 - 4(\xi r'-r)^2 + 4(r')^2(r'+2n) = 0, \label{eq:P4} 
\ee

\noindent where $r = T'$ and `prime' means the derivative w.r.t. the endpoint $\xi$, since for the case of single endpoint it turns into the equation (\ref{eq:P4}). The last term in the parentheses of eq.~(\ref{eq:Pr}), which has no analog in eq.~(\ref{eq:P4}), turns then to zero.
\par However, the equation (\ref{eq:Pr}) is not the only third-order PDE satisfied by $T$, see the others for the 2-endpoint case in the sections above. Thus, this situation that the gap probabilities for more than one spectral endpoint satisfy several different PDE is very general and applies for both single and coupled ensembles.

%\bibliography{UniUE}

\end{document}